\documentclass[aps,pra,amsmath,amssymb,showpacs,twocolumn]{revtex4}

\usepackage{amssymb}
\usepackage{amsmath}
\usepackage{amscd}
\usepackage{natbib}
\usepackage{graphicx}            

\newcommand{\be}{\begin{equation}}
\newcommand{\ee}{\end{equation}}

\renewcommand{\vec}[1]{{\mathbf #1}}

\begin{document}


\title{The Fermi-polaron in two dimensions: Importance of the two-body bound state}

 \author{Michael Klawunn and Alessio Recati}
 \affiliation{
 \mbox{INO-CNR BEC Center and Dipartimento di Fisica, Universit\`a di Trento, 38123 Povo, Italy}}

\date{\today}

\begin{abstract}
We investigate a single impurity interacting with a free two-dimensional atomic Fermi gas.
The interaction between the impurity and the gas is characterized by an arbitrary attractive short-range potential,
which, in two dimensions, always admits a two-particle bound state. 
We provide analytical expressions for the energy and the effective mass of the dressed impurity by including the two-body bound state,
which is crucial for strong interactions, in the integral equation for the effective interaction.
Using the same method we give also the results for the polaron parameters in one and three dimensions finding a pretty good agreement
with previous known results. Thus our relations can be used as a simple way to estimate the polaron parameters once the two-body bound state of the interaction potential is known.
\end{abstract}


\maketitle
\section{Introduction}

In the last years, thanks to the experimental capability of tuning 
the relative population and the interaction strength in atomic gases of different species,  highly imbalanced gases have been extensively studied.
In particular a lot of theoretical and experimental work has been devoted to atomic Fermi gases in two different hyperfine states (see, e.g., the recent review \cite{ChevyMora} and references therein).
The building block is the solution of the limiting case of a single impurity atom interacting via a short range potential with an ideal atomic Fermi gas.
Such a problem is not only relevant in the field of ultra-cold gases, since it is related to the more general one, the so-called impurity problem, which is present also in other area of physics. 
In our case the dressed impurity is called Fermi-polaron (or polaron) in analogy with electrons dressed by the bosonic (phonon) bath in a crystal.

Important quantities characterizing the polaron are: (i) its chemical potential, also called interaction energy or binding energy, i.e.,
the (negative) energy difference of the ground state with and without
the impurity atom at rest; (ii) its effective mass, i.e., the dressed parabolic dispersion relation at low impurity's momentum.
In three dimensions (3D) these parameters have been calculated in different ways, e.g., by means of variational ansatz  \cite{Alessio2007, Mol}, Monte-Carlo methods \cite{Prokofiev, PilatiGiorgini}, functional renormalisation group \cite{Schmidt2011} and they have been measured experimentally \cite{ExpPol}. 

The variational approach is known to give reasonable results also in the one-dimensional (1D) case \cite{Alessio2007,Combescot1D}, since it can be compared with the exact solution found by McGuire \cite{McGuire}.
Very recently the same approach has been used to study the two dimensional (2D) case, where it has been shown that in 2D its use can be questionable at least if using the same approximations as in 3D \cite{var2D-1,var2D-2}.

In two-dimensions, as well as in one dimension, an attractive interaction always allows for a two-body bound state. In the present work we solve the impurity problem including such a bound state explicitly in the integral equation for the effective interaction of the impurity with the Fermi gas. Within a number of approximations we can provide analytical expressions for the polaron parameters which agree quite well with the known results in one and three dimensions. In 2D we find an expression for the energy which interpolates between the correct/expected limiting values in the weakly and the strongly interacting regime.
Thanks to the recent experimental advances in realizing one- and two-dimensional strongly interacting Fermi gases \cite{Hulet,Koehl},
the impurity problem in reduced dimensionality has become relevant in the context of ultra-cold gases. 
 

In the next section we introduce the formalism and give the result for the energy and the effective mass of the two-dimensional Fermi-polaron problem.
In Sec. \ref{13D} we apply the method to the one- and three-dimensional case.

\section{Fermi-polaron in 2D}

%
It is known that Brueckner-Hartree-Fock theory, when applied to the Fermi-polaron problem in three dimension gives reasonable results \cite{Lipparini} (see also Sec. \ref{sec:3d}).
The basic equation from this theory is the Bethe-Goldstone integral equation
for the reaction matrix \cite{BG1957}, also called effective interaction.
(e.g. for the 2D electron gas \cite{Nagano1984}). 
The Bethe-Goldstone integral equation for the effective interaction between a particle in the bath with momentum $\vec{k_1}$ and 
the impurity atom with momentum $\vec{k_2}$ can be written as
\begin{align}\label{BG}
\begin{split}
&g(\vec{k_1},\vec{k_2},\vec{q})=V(\vec{q})+\int \frac{d\vec{k}}{(2\pi)^D}
V(|\vec{q}-\vec{k}|)\times\\
& \frac{\left(1-n_{\vec{k_1}+\vec{k}}\right) }
{\frac{k_1^2}{2m}+{\frac{k_2^2}{2m}}-\frac{(\vec{k_1}+\vec{k})^2}{2m}
-\frac{(\vec{k_2}-\vec{k})^2}{2m}}
g(\vec{k_1},\vec{k_2},\vec{k})
\end{split}
\end{align}
In Eq. (\ref{BG}) $\vec{q}$ is a transfer momentum, $V(\vec{q})$ is the Fourier transform
of the two-particle interaction potential, $n_{\vec{k}}$ the Fermi distribution function at
zero temperature.
The interaction energy or correlation energy follows then from the mean value of the effective interaction
$\epsilon_p= \left\langle g(\vec{k_1},\vec{k_2},\vec{q})\right\rangle$

For  $\vec{k_2}=0$ one gets the rest correlation energy $\epsilon_p^0$
of the polaron, and by expanding about this solution in $k_2^2$ one get its effective mass $m^*$ as usual
by the relation $E = \epsilon_p^0+k_2^2/2m^*$.  
We remind that in Eq. (\ref{BG}) for the effective interaction only ladder diagrams are 
summed and the Fermi sea limits the momenta in the intermediate states.

We consider an interaction characterized by an attractive short-range potential of arbitrary shape.
In three dimensions this potential can be approximated by a delta function,
and Eq. (\ref{BG}) coincides with the self-consistent equation obtained via single particle-hole variational ansatz \cite{Alessio2007,Lipparini}.
In two dimensions it is not clear, whether one can 
use a delta function pseudo-potential, (see e.g. \cite{Yang2008}), hence, a solution of Eq. (\ref{BG}) obtained in the same way
as in 3D is questionable. As usual in order to treat properly the 2-body problem we write Eq. (\ref{BG})
by expressing $V$ in terms of the two-particle scattering amplitude $f$ \cite{Abrikosov-book}
\begin{align}
\begin{split}\label{BG-polaron-f1}
&g(\vec{k_1},\vec{k_2},\vec{q})=f(\vec{k_1}-\vec{k_2},\vec{q})+\phantom{\int}\\
\int\!\! \frac{d\vec{k}}{(2\pi)^D}
& \frac{f(\vec{k_1}-\vec{k_2}, \vec{q}-\vec{k} ) \left(-n_{\vec{k_1}+\vec{k}}\right) }
{\frac{k_1^2}{2m}+{\frac{k_2^2}{2m}}-\frac{(\vec{k_1}+\vec{k})^2}{2m}
-\frac{(\vec{k_2}-\vec{k})^2}{2m}}
g(\vec{k_1},\vec{k_2},\vec{k}),
\end{split}
\end{align}
where $f(\vec{k_1}-\vec{k_2},\vec{q})$ is the off-shell scattering amplitude \cite{note1}.
Note, that this equation is already renormalized with repect to ultraviolet divergencies.
 
For short-range potentials the exchange momentum $\vec{q}$ is small and the main contribution
from the integral comes from small values of $\vec{k}$, thus we can
approximate the off-shell scattering amplitude
by the on-shell scattering amplitude $f(\vec{k_1}-\vec{k_2})$. Then the 
effective interaction does not depend on the exchange momentum
and for the impurity at rest ($k_2=0$) Eq. (\ref{BG-polaron-f1}) reduces to
\begin{align}\label{g}
\frac{1}{g(\vec{k_1})}\approx \left[ \frac{1}{f(\vec{k_1})}-\int_{p<k_F} \frac{d\vec{p}}{(2\pi)^2}
 \frac{m}
{\vec{p} \left(\vec{p}- \vec{k_1} \right)}\right] 
\end{align}

We assume that the  finite range $R$ of the attractive interaction potential is the shortest
length scale in the system. In particular for $k_F R \ll1$ the $s$-wave scattering amplitude 
reads (see e.g. \cite{Randeria1990})
\be f(k)^{-1}=-[\,\ln(k^2/m|\epsilon_b|)-i\pi\,]m/4\pi
\label{fk}\ee
where $\epsilon_b$ is the binding energy of the two-particle bound state,
which in 2D (differently from 3D) is always present \cite{eb}. 

Solving Eq.(\ref{g}) for the weakly interacting case $|\epsilon_b| \ll \epsilon_F$,
where $\epsilon_F=\hbar^2 k_F^2/(2m)$ is the Fermi energy,
we find the interaction
$g(k_1)=-4\pi / [m \ln(2\epsilon_F/|\epsilon_b|)]$ from which we get the mean field energy
 $\epsilon_p^0=-2\epsilon_F/ [m \ln(2\epsilon_F/|\epsilon_b|)]$, which is obviously in agreement with the result
 found in the weakly interacting regime using the single particle-hole variational ansatz \cite{var2D-1}.

Until now, deriving Eq. (\ref{g}) from the Bethe-Goldstone Equation, we have neglected the two-particle bound state.
In order to take it into account we go back to Eq. (\ref{BG-polaron-f1}).
We rewrite the initial energy of excitation processes appearing in the denominator as
$\frac{k_1^2}{2m}+{\frac{k_2^2}{2m}}=\frac{k_r^2}{m}+{\frac{P^2}{4m}}$,
with relative momentum $\vec{k_r}$ and centor-of-mass momentum $\vec{P}$. 
Further, we remind that the scattering amplitude $f(\vec{k_r})$ depends on the relative momentum only.
It is well known that in the case where the majority particle with $k_1$ and the impurity with $k_2$
form a two-body bound state, their relative momentum $k_r$ is purely imaginary
with $\frac{k_r^2}{m}=\epsilon_b<0$ being the binding energy.
For the calculation of the interaction energy we assume the impurity
to be at rest and $P=0$.
Then the effective interaction in ladder approximation obeys instead of Eq. (\ref{g})
\begin{align}\label{g-bound}
\frac{1}{g(\epsilon_b)} \approx \frac{1}{f\left( k_r\right) } - \int_{k<k_F} \frac{d\vec{k}}{(2\pi)^2}
 \left(|\epsilon_b|+\frac{k^2}{m}\right)^{-1}. 
\end{align}
Let us notice that a very similar equation is found in \cite{Vagov2007} for the vertex function 
in the presence of a two particle bound-state, where the molecular propagator
is expressed by the two-particle scattering amplitude at the vacuum energy
of the molecule.

\begin{figure}[ht]
\begin{center}
\includegraphics[width=0.44\textwidth,angle=0]{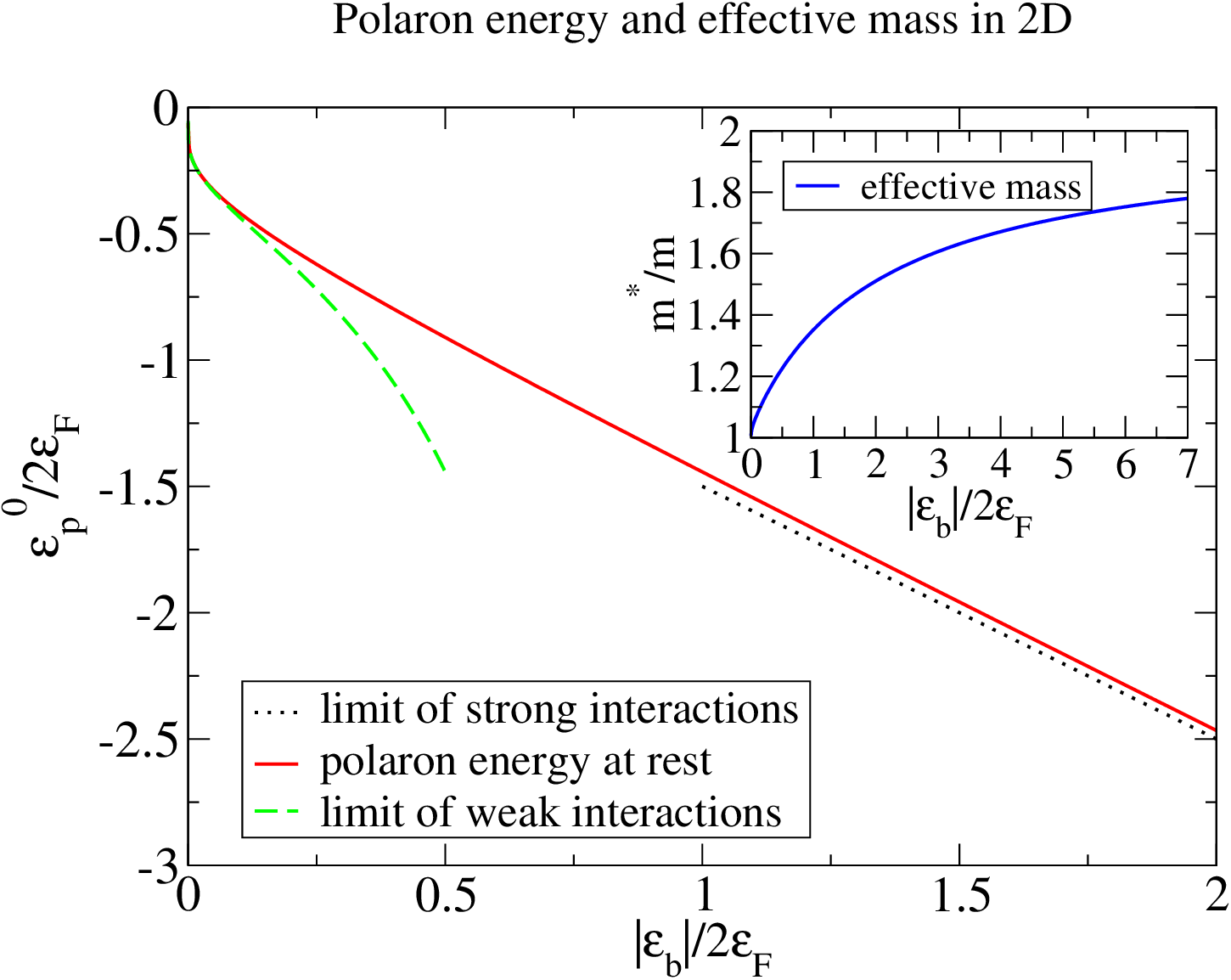}
\caption{Main: polaron energy as a function of the two-body binding energy $\epsilon_b$ as given by Eq. (\ref{E_int_bound}) (solid-red line). For completeness we report also the weakly interacting (dashed-green line) and the strongly interacting (dotted-black line) results (see text).
Inset: ratio $m^*/m$ between the effective and the bare mass as given by Eq. (\ref{ms-2d}).}
\label{fig:E_int}
\end{center}
\end{figure}

At the momentum corresponding to the bound state the scattering amplitude has a pole $f\left( k_r=\sqrt{m\epsilon_b}\right)^{-1}=0$
and the interaction energy is given by
\begin{align}\label{E_int_bound}
\epsilon_p^0 \approx n g(\epsilon_b)=\frac{-2\epsilon_F}{\ln\left[1+\frac{2\epsilon_F}{|\epsilon_b|} \right]},
\end{align}
and it is shown in Fig. \ref{fig:E_int}.

In the limit of weak interactions, i.e.,
$|\epsilon_b| \ll \epsilon_F$, the polaron energy reduces to
$\epsilon_p\approx -2\epsilon_F/ \ln(2\epsilon_F/|\epsilon_b|)$,
which coincides with the one obtained from Eq. (\ref{g}) or using the variational ansatz \cite{var2D-1}, 
where the two-particle bound state is not included. 
In the opposite limit, $|\epsilon_b| \gg \epsilon_F$, 
Eq. (\ref{E_int_bound}) yields 
$\epsilon_p= -|\epsilon_b|-\epsilon_F+o(\epsilon_F/|\epsilon_b|)$.
This is the expected result, because in this regime
one atom of the majority is
strongly bound to the impurity with $-|\epsilon_b|$ and thus it has to be removed from the Fermi-sea
leading to the first correction $-\epsilon_F$. 
Thus Eq. (\ref{E_int_bound}) smoothly interpolates between these two
limits and it provides a good approximation for the interaction energy of the 2D polaron
in all regimes.
Our results are in good agreement with recent preliminary Monte-Carlo calculations \cite{Bertaina-private}.  
We remind that Eq. (\ref{E_int_bound}) is valid for all attractive potentials with $s$-wave scattering amplitude of
logarithmic form Eq. (\ref{fk}).

\subsection{Effective mass}\label{sec:meff}

In this section we study the effect of the interaction on the motion of the impurity.
As already mentioned its energy can be expanded for small momentum $k_2$ as $E=\epsilon_p^0+k_2^2/2m^*$,
where we define the effective mass of the impurity as
\begin{align}\label{def-ms}
\frac{1}{m^*}=\frac{1}{m}+\frac{1}{m}\frac{d \epsilon_p(\vec{k_2})}{d(k_2^2/2)}|_{k_2=0}
\end{align}
and $\epsilon_p(k_2) =n g(\epsilon_b,k_2)$, with the effective interaction $g$ calculated again including the two-body bound state.
In particular the initial energy of excitation processes
is $-|\epsilon_b|+k_2^2/4m$ and instead of Eq. (\ref{g-bound}) we obtain 
\begin{align}\label{g-bound-k2}
\frac{1}{g(\epsilon_b,\vec{k_2})} \approx \frac{1}{f\left( k_r\right) }
 - \int_{0}^{k_F}\!\! \frac{d\vec{k}}{(2\pi)^2}
 \frac{1}{|\epsilon_b| +e(k_2,k)}
\end{align}
where $ e(k_2,k)=\frac{k_2^2}{4m}+ \frac{k_2 k \cos\phi}{m}+\frac{k^2}{m}$.
The ratio between the bare and the effective mass of the impurity atom in two dimensions within our approximation reads
\begin{align}\label{ms-2d}
\frac{m}{m^*}=
1-\frac{1}{2}\left( \frac{\epsilon_p^0}{2\epsilon_F}\right)^2 \left( 1+\frac{|\epsilon_b|}{2\epsilon_F}\right) ^{-2}. 
\end{align}
As shown in Fig. \ref{fig:E_int} the effective mass $m^*$ obtained from the previous equation has the expected behavior: for small interactions it
is close to the bare mass value $m$ and for large interactions it approaches the molecular mass value $2m$.

\section{Fermi-polaron parameters in 1D and 3D}
\label{13D}

For a  two-dimensional system our approach seems to give quite reasonable results and provides analytical expressions for the polaron's parameters. 
In the present section we apply our approach to the one- and three-dimensional case. 
The simple expressions we find are in reasonable agreement with the known results.

\subsection{The Fermi-polaron in one dimension}

In one dimension the Fermi-polaron problem admits an exact solution \cite{McGuire} and
the interaction energy reads  
\begin{align}\label{E_int-1D-exact}
\frac{\epsilon_p^0}{2\epsilon_F}=-\frac{1}{\pi}\left[ 
y+\frac{\pi}{2} y^2 + (1+y^2) \arctan(y) \right], 
\end{align}
where $y=\sqrt{|\epsilon_b| / (2\epsilon_F)}$. Again
$\epsilon_b$ is the binding energy of the lowest two-body bound
state. 

\begin{figure}[ht]
\begin{center}
\includegraphics[width=0.44\textwidth,angle=0]{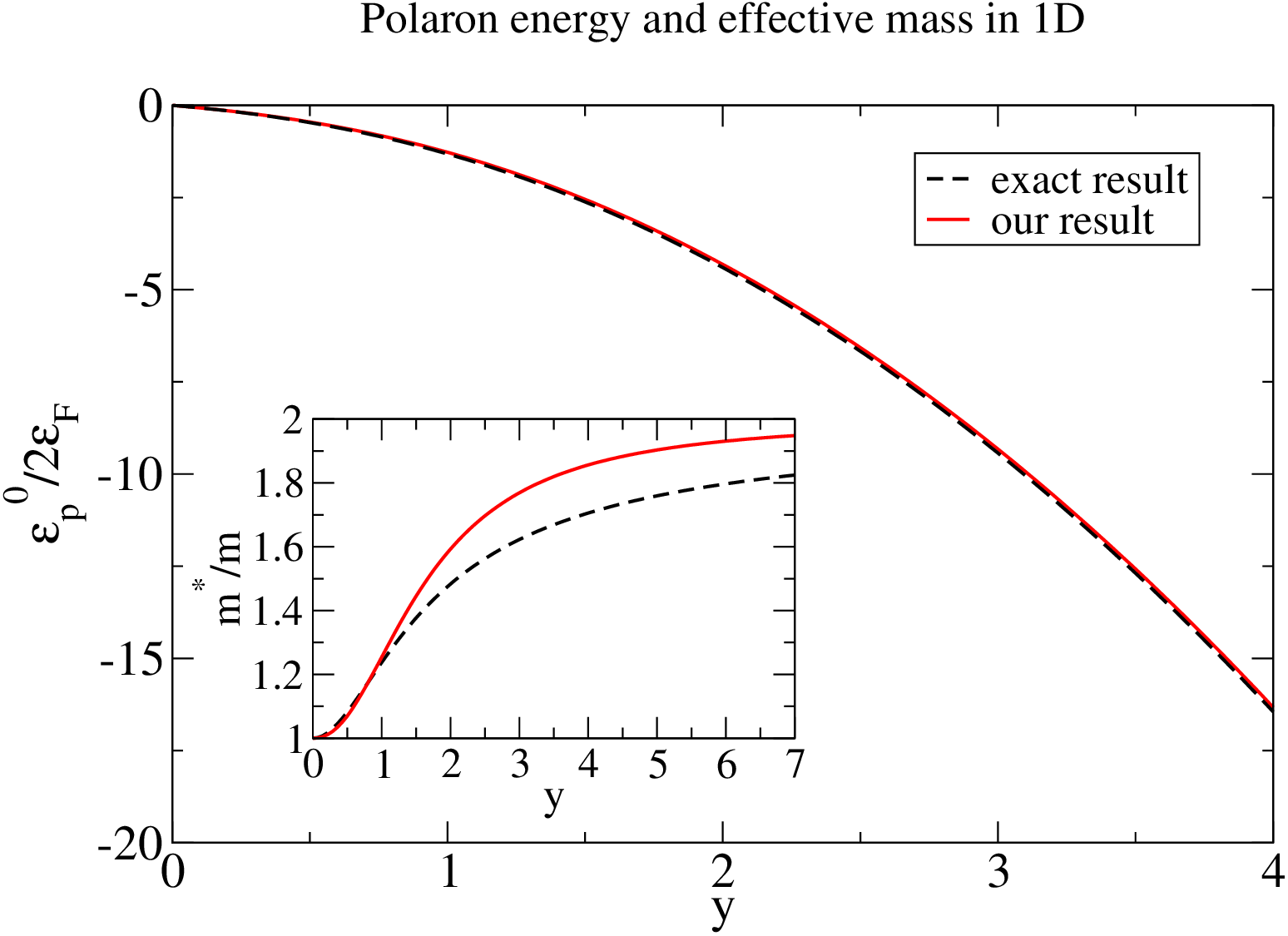}
\caption{Interaction energy (main panel) and effective mass (inset) as a function of the two-body binding energy $\epsilon_b$ in 1D. 
The approximate result Eq. (\ref{E_int-1D}) (solid-red line) is compared with the exact expression Eq. (\ref{E_int-1D-exact}) given by McGuire \cite{McGuire}. 
}
\label{fig:E_int-1D}
\end{center}
\end{figure}

When applied to 1D Eq. (\ref{g-bound}) gives for the polaron energy
\begin{align}\label{E_int-1D}
\frac{\epsilon_p^0}{2\epsilon_F} \approx -\frac{y}{\arctan(\frac{1}{y})},
\end{align}
which we compare against the exact result Eq. (\ref{E_int-1D-exact}) in Fig. \ref{fig:E_int-1D}.
The agreement looks pretty good, although in the strongly interacting case, we get $-|\epsilon_b|-2/3\epsilon_F$ instead of
$-|\epsilon_b|-\epsilon_F$.
However our results are closer to the exact solution than the one obtained in with the single particle-hole variational ansatz (see e.g., \cite{Combescot1D}).

\subsection{The Fermi-polaron in three dimensions.}
\label{sec:3d}

\begin{figure}[ht]
\begin{center}
\includegraphics[width=0.45\textwidth,angle=0]{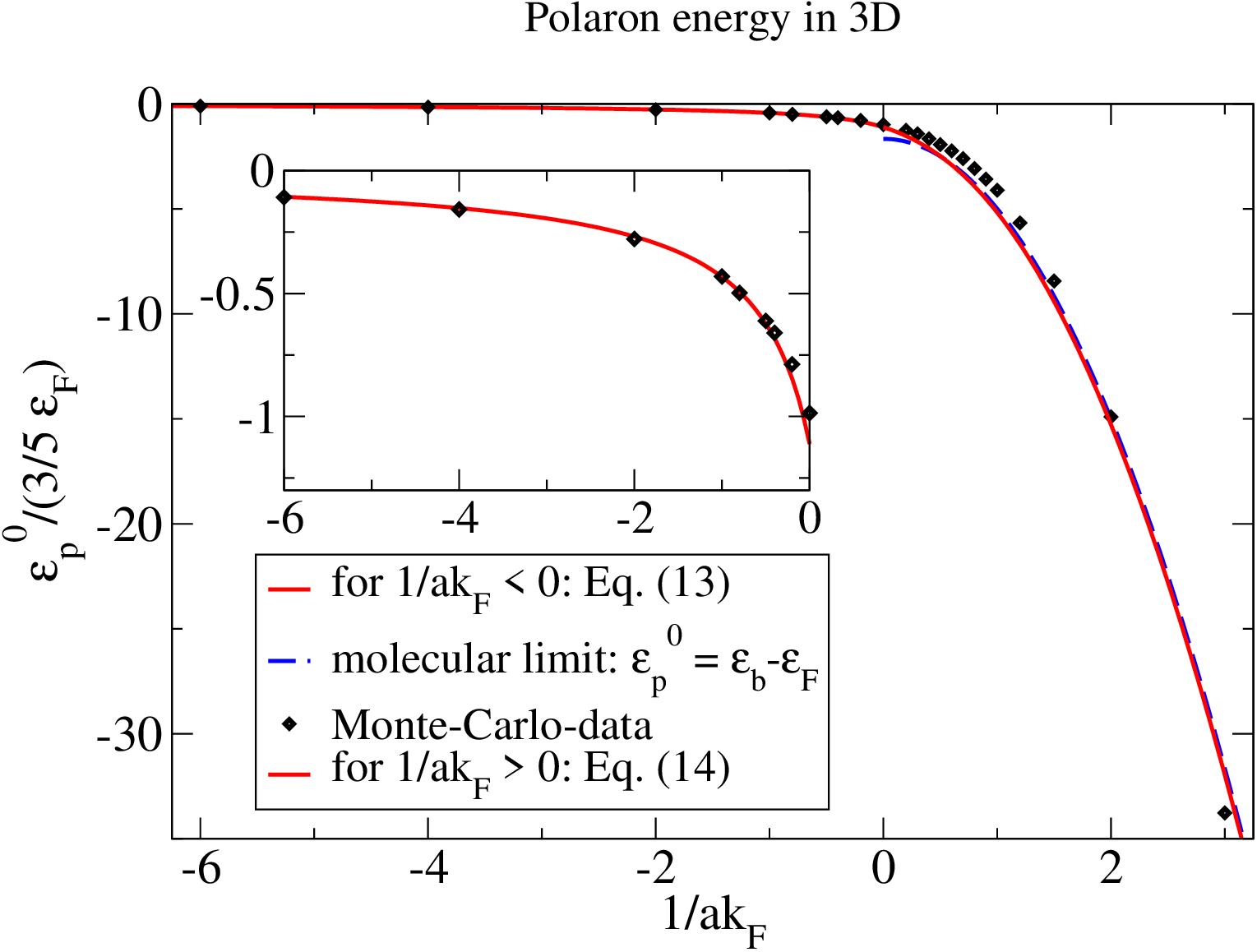}
\caption{Interaction energy as a function of the inverse 3D scattering length $1/(k_F a)$ (red line)
in comparison with the results obtained from Monte-Carlo calculations \cite{PilatiGiorgini}
in the polaron regime and at unitarity (black diamonds). In the molecule regime ($|\epsilon_b| \gg \epsilon_F$)
we compare the expected result $\epsilon_{int}^0=-|\epsilon_b|-\epsilon_F$ (blue line).
Inset: zoom on the negative axis.} 
\label{fig:E_int-3d}
\end{center}
\end{figure}

In a three-dimensional geometry when no two-body bound state is present, i.e., for negative $s$-wave scattering length ($a<0$), we can use Eq. (\ref{g}) -- the usual Brueckner-Hartree-Fock theory --  with scattering amplitude 
$4\pi f(\vec{k_1})^{-1}/m=a^{-1}+i|\vec{k_1}|$. The main contribution to the effective interaction is
$g(k_1=0)$ and thus we can write an approximated expression for the polaron energy as
\begin{align}\label{polaron-3d}
\frac{\epsilon_{p}^0}{\epsilon_F}\approx-\frac{2}{3\left( 1-\frac{\pi}{2}\frac{1}{k_F a}\right)}.
\end{align}
Note, that adding self-consistency to the Bethe-Goldstone equation by changing the initial energy of excitation processes
$\frac{k_1^2}{2m}+{\frac{k_2^2}{2m}} \to \epsilon_p^0+\frac{k_1^2}{2m}+{\frac{k_2^2}{2m}}$ \cite{Lipparini}
increases the accuracy of the results in the polaron regime and yields the same equations
as in \cite{Alessio2007}. 

In the molecular regime the potential admits a two-body bound-state ($a>0$) with binding energy
$\epsilon_b=-1/(ma^2)$.  From the 3D version of Eq. (\ref{g-bound}) we obtain the interaction energy 
of the impurity, which reads
\begin{align}\label{bound-3d}
\frac{\epsilon_{p}^0}{\epsilon_F}\approx-\frac{2}{3\left( 1-\sqrt{\frac{|\epsilon_b|}{2\epsilon_F}}
\arctan\left( {\sqrt{\frac{2\epsilon_F}{|\epsilon_b|}}}\right) \right) },
\end{align}
with $\sqrt{|\epsilon_b|/(2\epsilon_F)}=1/(k_F a)$.
For large binding energy $|\epsilon_b| \gg \epsilon_F$ (or $k_F a\ll 1$), 
one gets $\epsilon_{int}^0=-|\epsilon_b| - 6\epsilon_F/5$,
which is larger than the expected result by $-1/5$ $\epsilon_F$.
In Fig. \ref{fig:E_int-3d} we show the comparison between the previous simple expression and the results obtained from Monte-Carlo calculations \cite{PilatiGiorgini}.
Although the agreement is good it is worse, as already mentioned, than the results obtained from the
variational approach in both the polaron \cite{Alessio2007} and the molecular regime \cite{Mol} as well as the results
obtained from the functional renormalization group \cite{Schmidt2011}. 

\subsection{Remarks on the effective mass in 1D and 3D}

The effective mass of the impurity is given by Eq. (\ref{ms-2d}) in any dimension. 

In the inset of Fig. \ref{fig:E_int-1D} we compare our result (Eq. \ref{ms-2d}) with the exact one found by McGuire in  \cite{McGuire} which 
reads
\begin{align}\label{E_int-1D-ms}
\frac{m^*}{m}&=
\frac{\left( 1+\frac{2}{\pi}\arctan{y}\right) ^2}{1+\frac{2}{\pi}\left( \arctan{y}+\frac{y}{1+y^2}\right) },
\end{align}
where $y=\sqrt{|\epsilon_b| / (2\epsilon_F)}$ is defined as in Eq. (\ref{E_int-1D-exact}).
Again the agreement is reasonable and better than the single particle-hole variational ansatz \cite{Alessio2007,Combescot1D}.

In 3D the situation is more involved since a maximum (quite larger than $2 m$) for the effective mass has been found when the nature of the impurity changes from a fermionic quasi-particle to a bosonic quasi-particle \cite{Prokofiev}.  In our approximation this maximum cannot be found. 
In order to find this maximum one has to take three-particle scatteirng into account which is beyond the scope of this paper.

\section{Conclusion}

In conclusion, we have investigated the problem
of an impurity atom interacting with a non-interacting Fermi-gas
in a two-dimensional geometry. We consider a short range, attractive potential, which
implies the presence of a two-bound state for any interaction strength. 
We have calculated the interaction energy and the effective mass of the impurity
by including the bound-state in the Bethe-Goldstone integral equation for
the effective interaction. We were able to obtain simple analytical expressions
which give reasonable results in 2D, interpolating between the weakly and the strongly interacting 
regime (see Fig. \ref{fig:E_int}). Moreover when applied to the three- and one-dimensional case
our polaron parameters well compare with most of the known results (see Sec. \ref{13D}). 
Thus our expressions can be used to estimate the dressed impurity's parameters in a simple way once the 
two-body bound state is known.
Our analysis shows how important the two-body bound state is for the polaron problem.

Finally let us stress that we do not address the question of whether there exists a polaron-to-molecule transition in two-dimension as debated in \cite{var2D-1,var2D-2}. The problem is still open and it could happen that the system behaves similarly to the one-dimensional case where there is not such a transition. Such a question is clearly relevant for the possible low-temperature phases of a two-dimensional polarized Fermi gas, whose balanced version has been recently experimentally realized \cite{Koehl}.

\acknowledgments
Useful discussions with Georg Bruun and Alexander Pikovski are acknowledged. 
We also thank G. Bertaina and S. Giorgini for the opportunity 
to compare our calculations with the unpublished results of a Monte-Carlo calculation.
This work has been supported by ERC through the QGBE grant.


\begin{thebibliography}{99}

\bibitem{ChevyMora}
F. Chevy and C. Mora, Rep. on Prog. in Phys. {\bf 73},   112401 (2010).

\bibitem{Alessio2007}
R. Combescot, A. Recati, C. Lobo and F. Chevy, Phys. Rev. Lett. {\bf 98}, 180402 (2007).
 
\bibitem{Mol} 
C. Mora and F. Chevy, Phys. Rev. A {\bf 80}, 033607 (2009);
M. Punk, P. T. Dumitrescu, and W. Zwerger, Phys. Rev. A {\bf 80}, 053605  (2009);
R. Combescot, S. Giraud, and X. Leyronas, Eur. Phys. Lett. {\bf 88}, 60007 (2009).

\bibitem{Prokofiev} 
N. Prokof'ev and B. Svistunov, Phys. Rev. B {\bf 77}, 020408 (2008).

\bibitem{PilatiGiorgini}
S. Pilati and S. Giorgini, Phys. Rev. Lett. {\bf 100}, 030401 (2008).

\bibitem{Schmidt2011}
R. Schmidt and T. Enss,
eprint: arXiv: 1104.1379.

\bibitem{ExpPol}
A. Schirotzek, C.-H. Wu, A. Sommer and M. W. Zwierlein,
Phys. Rev. Lett. {\bf 102}, 230402 (2009);
S. Nascimbene, N. Navon, K. J. Jiang, L. Tarruell, M. Teichmann, J. McKeever, F. Chevy, and C. Salomon, Phys. Rev. Lett. {\bf 103}, 170402 (2009);
N. Navon, S. Nascimbene, F. Chevy, C. Salomon, Science {\bf 328}, 5979 (2010).

\bibitem{Combescot1D}
S. Giraud and R. Combescot, Phys. Rev. A {\bf 79}, 043615 (2009).

\bibitem{McGuire}
J. B. McGuire, J. Math. Phys. (N.Y.) 7, 123 (1966).

\bibitem{var2D-1}
S. Z\"ollner, G. Bruun and C. J. Pethick,
Phys. Rev. A {\bf 83}, 021603 (2011).

\bibitem{var2D-2}
M. Parish, 
Phys. Rev. A {\bf 83}, 051603 (2011).

\bibitem{Hulet}
Y.-an Liao, A. S. C. Rittner, T. Paprotta, W. Li, G. B. Partridge,	 R. G. Hulet, S. K. Baur, and E. J. Mueller,
Nature {\bf 467}, 567 (2010).

\bibitem{Koehl}
B. Fr\"ohlich, M. Feld, E. Vogt, M. Koschorreck, W. Zwerger, M. K\"ohl,
Phys. Rev. Lett. {\bf 106}, 105301 (2011).

\bibitem{Lipparini}
E. Lipparini,
{\em Modern many particle physics (2nd edition)},
World Scientific 2008.


\bibitem{BG1957}
H. A. Bethe and J. Goldstone,
Proc. Roy. Soc. A {\bf 238}, 551 (1957). 

\bibitem{Nagano1984}
S. Nagano, K. S. Singwi and S. Ohnishi,
Phys. Rev. B {\bf 29}, 1209 (1984).

\bibitem{Yang2008}
C. N. Yang,
Euro. Phys. Lett. {\bf 84}, 40001 (2008).


\bibitem{Abrikosov-book}
A. A. Abrikosov, L. P. Gorkov, I. E. Dzyaloshinski,
{\em Methods of quantum field theory in statistical physics,}
Prentice-Hall, 1963

\bibitem{note1}
Our notation of the off-shell scattering amplitude differs slightly from
the usual one \cite{Abrikosov-book}:
$f(\vec{k_1},\vec{q}) \equiv f(\vec{k_1},\vec{k_1}+\vec{q})$
Thus we can approximate it by the on-shell amplitude $f(\vec{k_1})$
if the second argument is small.

\bibitem{Randeria1990}
M. Randeria, J.-M. Duan and L.-Y. Shieh,
Phys.~Rev.~B {\bf 41}, 327 (1990);

\bibitem{eb}
The relation between the binding energy and the potential $V(\bf{r})$ in two-dimensions has been recently 
discussed in detail in \cite{Michael2D}
 
\bibitem{Michael2D}
M. Klawunn, A. Pikovski and L. Santos,
Phys. Rev. A {\bf 82}, 044701 (2010).


\bibitem{Vagov2007}
A. Vagov, H. Schomerus and A. Shanenko,
Phys. Rev. B {\bf 76}, 214513 (2007).

\bibitem{Bertaina-private}
G. Bertaina and S. Giorgini, private communication.

\end{thebibliography}
\end{document}